\documentstyle[12pt]{article}

\def\ut#1{\rlap{\lower1ex\hbox{$\sim$}}#1{}}

\begin{document}
\begin{center}
   \vskip 2em
  {\LARGE Self-dual gravity with topological terms}
   \vskip 3em
  {\large Merced Montesinos \footnote{E-mail: merced@fis.cinvestav.mx}\\[2em]}
\em{
Departamento de F\'{\i}sica \\
Centro de Investigaci\'on y de Estudios Avanzados del I.P.N. \\
Av. I.P.N. 2508, 07000, Ciudad de M\'exico, M\'exico \\[1em] }

\end{center}
 \vskip 1em
\begin{abstract}
The canonical analysis of the (anti-) self-dual action for gravity
supplemented with the (anti-) self-dual Pontrjagin term is carried out. The
effect of the topological term is to add a `magnetic' term to the original
momentum variable associated with the self-dual action leaving the Ashtekar
connection unmodified. In the new variables, the Gauss constraint retains its
form, while both vector and Hamiltonian constraints are modified. This shows,
the contribution of the Euler and Pontrjagin terms is not the same as that
coming from the term associated with the Barbero-Immirzi parameter, and thus
the analogy between the $\theta$-angle in Yang-Mills theory and the
Barbero-Immirzi parameter of gravity is not appropriate.
\end{abstract}
\date{\today}
\vskip 1em
PACS: 04.60.Ds
\vskip 3em

\section{Introduction}
In the first order formalism gravity supplemented with topological terms is
given by the action
\begin{eqnarray}
S [ e , \omega ] & = &
+ \alpha_1 \; \int \ast ( e^I \wedge e^J ) \wedge R_{IJ} (\omega)
+  \alpha_2  \; \int e^I \wedge e^J  \wedge R_{IJ} (\omega) \nonumber\\
& & + \alpha_3 \int R^{IJ} (\omega) \wedge R_{IJ} (\omega) +
\alpha_4 \int {\ast R}^{IJ} (\omega) \wedge R_{IJ} (\omega)  \,.
\label{action}
\end{eqnarray}
The first term in (\ref{action}) is the Hilbert-Palatini action, the second
one is proportional to the first Bianchi identities when there is no torsion
and thus vanishes `on shell,' namely, in a second order formalism. Third and
fourth terms  are the Pontrjagin and Euler terms, respectively. $e^I$ is a
non-degenerate inverse tetrad frame with Lorentz indices $I$, $J = 0,1,2,3,$;
raised and lowered with the Lorentz metric $\eta^{IJ}$. The spacetime
signature is $\eta_{IJ}=\mbox{diag} (-1,+1,+1,+1)$. $\omega_I \,^J$ is a
Lorentz connection 1-form and $R_{IJ} (\omega) = \frac{1}{2} R_{\mu\nu IJ}
(\omega) dx^{\mu} \wedge dx^{\nu}$ is its curvature, $R_{\mu\nu I}\,^J =
\partial_{\mu} \omega_{\nu I}\,^J - \partial_{\nu}
\omega_{\mu I}\,^J + \omega_{\mu I}\,^K \omega_{\nu K}\,^J -
\omega_{\nu I}\,^K \omega_{\mu K}\,^J$. The definition of the dual operator
is ${\ast T}^{IJ} = \frac{1}{2} \epsilon^{IJ}\,_{KL} T^{KL}$ with
$\epsilon_{0123} = +1$.

As far as I know, the full canonical analysis of action (\ref{action}) has
not been carried out. When $\alpha_2 = \alpha_3 = \alpha_4 =0$, action
(\ref{action}) reduces to the standard Hilbert-Palatini action whose
canonical analysis is already reported in the literature \cite{Lect}. There,
the canonical variables are the 3-dimensional extrinsic curvature $K^i_a$ and
the densitizied inverse triad field ${\widetilde E}^a_i$ when the time gauge
is chosen. On the other hand, (anti-) self-dual gravity is obtained when
$\alpha_2 =\epsilon i \alpha_1$, and $\alpha_3 = \alpha_4 =0$ with
$\epsilon=+$ and $\epsilon=-$ for the self-dual and anti-self-dual actions,
respectively \cite{Pleb77,Sam87,Jacob88,Capo91a}. The canonical formalism of
the (anti-) self-dual action leads to the phase space variables introduced by
Ashtekar \cite{Abhay}. These phase space variables are a complex $SU(2)$
connection and the densitizied inverse triad field when the time gauge is
fixed. The canonical formalism of the self-dual action can be carried out
without fixing the gauge, which implies modifications on both the
configuration variable and its momentum \cite{Mon01}. When $\alpha_3 =
\alpha_4 =0$ and $\alpha_2$ is a non vanishing real parameter, action
(\ref{action}) reduces to the action studied by Holst \cite{Holst}, whose
canonical formalism leads to the phase space variables introduced by Barbero;
which were initially found via a canonical transformation from the phase
space variables associated with the Hilbert-Palatini action \cite{Bar}. One
of the advantages of Barbero variables is that they are real for Lorentzian
gravity and their use has been crucial for the development of the quantum
theory, known as loop quantum gravity [see, for instance, Ref.\cite{loop}].

The parameter $\alpha_2$ gives rise to the Barbero-Immirzi parameter. Both
parameters $\alpha_1$ and $\alpha_2$ enter in the spectra of geometric
operators \cite{Giorgio}. The usual choice is $\alpha_1 =\frac{c^3}{16\pi G}$
leaving $\alpha_2$ arbitrary. Up to now, Barbero-Immirzi parameter has not
been fixed either by using fundamental principles nor by experimental means.
It has been shown by Rovelli-Thiemann that the quantizations coming from
various values for the Barbero-Immirzi parameter are inequivalent \cite{Rov},
which is not a particular fact of field theory and thus of gravity. In fact,
the same happens even for systems with a finite number of degrees of freedom,
the reason being that the group of canonical transformations is not
isomorphic to the group of unitary transformations, so it is natural that
systems related by canonical transformations have, in general, inequivalent
quantum theories. In addition, it has been argued by Gambini-Obreg\'on-Pullin
that the Barbero-Immirzi parameter is very similar to the $\theta$ ambiguity
present in Yang-Mills theories \cite{Gam} in the sense that both are
ambiguities of their respective quantum theories. Nevertheless, in Yang-Mills
theories the term associated with the $\theta$ angle corresponds with the
Pontrjagin term associated with the Faraday tensor while in gravity, as has
been noted, the term in the action (\ref{action}) which gives rise to the
Barbero-Immirzi parameter is the second term in (\ref{action}) which is
related to the first Bianchi identities when there is no torsion. Therefore,
the origin of both parameters, Barbero-Immirzi and $\theta$ angle, in both
theories is quite different in their respective actions.

Nevertheless, the question remains, how do Euler and Pontrjagin terms in
(\ref{action}) contribute in the canonical formalism of general relativity?
This is the issue addressed in this work. More precisely, the present
canonical analysis is restricted to the (anti-) self-dual case. In spite of
this, from the present analysis it will become clear that the contribution of
the second term in (\ref{action}), related to the Barbero-Immirzi parameter,
is completely different to the contribution of the third and fourth terms of
(\ref{action}). Actually, while the inclusion of the second term in
(\ref{action}) allows for the introduction of the Ashtekar-Barbero connection
$A^i_a$ instead of the extrinsic curvature $K^i_a$ as the gravitational
configuration variable, the effect of the third and fourth terms in
(\ref{action}) {\it does} add an extra piece to the expression for the
(anti-) self-dual momentum variable, leaving the Ashtekar connection
unmodified.

\section{Self-dual gravity}
Now follows a brief summary of the $3+1$ decomposition of the (anti-)
self-dual action for completeness reasons. As already mentioned, self-dual
($\epsilon=+$) and anti-self-dual ($\epsilon=-$) gravity are obtained by
setting in (\ref{action}) the parameters equal to $\alpha_2 = \epsilon i
\alpha_1$, $\alpha_3 =0$, $\alpha_4 =0$, from which the (anti-) self-dual
action follows:
\begin{eqnarray}
S [ e , \omega ] & = &
\alpha_1 \; \int \ast ( e^I \wedge e^J ) \wedge R_{IJ} (\omega)
+  \epsilon i \alpha_1  \; \int e^I \wedge e^J  \wedge R_{IJ} (\omega) \, ,
\label{SD}
\end{eqnarray}
which can be rewritten as
\begin{eqnarray}
S [e, {}^{(\epsilon)} \omega ] & = & \int \left [ 2 \epsilon i \alpha_1 \,\,
{}^{(\epsilon)} (e^I \wedge e^J) \wedge {^{(\epsilon)} R}_{IJ} \right ]\, ,
\label{SDII}
\end{eqnarray}
where ${}^{(+)} \omega$ and ${}^{(-)} \omega$ correspond with the self-dual
and anti-self-dual connections, respectively. By choosing the `time gauge,'
namely, $e^0= N dx^0$, $e^i = E^i_a N^a dx^0 + E^i_a dx^a$ where $a=1,2,3$
denotes space indices, action (\ref{SDII}) becomes
\begin{eqnarray}
S [A^i_a ,{\widetilde \Pi}^a_i , \ut{\lambda} , \lambda^a ,
\lambda^i ] & = & \int dx^0 dx^3 \left [ {\dot A}^i_a {\widetilde \Pi}^a_i -
(\ut{\lambda} {\widetilde {\widetilde H}} +
\lambda^a {\widetilde V}_a + \lambda^i {\widetilde G}_i )
\right ] \nonumber\\
& & + \int dx^0 \int dx^3 \partial_a ({\widetilde \Pi}^a_i \lambda^i) \, ,
\label{split}
\end{eqnarray}
where the dependence of the phase space variables and the Lagrange
multipliers on the initial Lagrangian variables is
\begin{eqnarray}
{\widetilde \Pi}^a_i & := & -2 \epsilon i \alpha_1 E E^a_i \, , \nonumber\\
A^i_a & := & \Gamma_{a}^i - \epsilon i \omega_{a {\hat 0}}\,^i
\, , \quad \Gamma^i_a = -\frac{1}{2} \epsilon^i\,_{jk} \omega_a\,^{jk}
\, , \nonumber\\
\ut{\lambda} & := & -\frac{1}{4 \alpha_1}\frac{N}{E} \, , \nonumber\\
\lambda^a & := & N^a \, , \nonumber\\
\lambda^i & := & - A^i_0 = - (\Gamma_0^i -
\epsilon i \omega_{0 {\hat 0}}\,^i )
\, , \quad \Gamma^i_0 = -\frac{1}{2} \epsilon^i\,_{jk} \omega_0\,^{jk} \, ,
\label{DEF}
\end{eqnarray}
and the first class constraints have the form
\begin{eqnarray}
{\widetilde{\widetilde H}} & := &
\epsilon^{ijk} {\widetilde \Pi}^a_i {\widetilde \Pi}^b_j
F_{abk}\ , \nonumber\\
{\widetilde V}_a & := & {\widetilde \Pi}^b_i F_{ab}\,^i \ , \nonumber\\
{\widetilde G}_i& := & {\cal D}_a  {\widetilde \Pi}^a_i =
\partial_a {\widetilde \Pi}^a_i + \epsilon_{ij}\,^k A^j_a
{\widetilde \Pi}^a_k \ , \label{CI}
\end{eqnarray}
with $F_{ab}\,^i = \partial_a A^i_b -\partial_b A^i_a +
\epsilon^i\,_{jk} A^j_a A^k_b$. When the `time gauge' is not
chosen, the definition of the phase space variables is modified.
In particular, the momentum ${\widetilde \Pi}^a_i$ is not just
proportional to the densitizied inverse triad field and
$\Gamma^i_a$ is not just the 3-dimensional spin connection if
contact with the second order formalism is required \cite{Mon01}.

Thus, in the Hamiltonian formulation of (anti-) self-dual gravity
is present the situation found in Yang-Mills and Maxwell theories.
In Maxwell theory, the Lagrangian action depends on a
4-dimensional connection $A_{\mu}$ and the action is fully
gauge-invariant under the gauge transformation $A_{\mu}
\rightarrow {A'}_{\mu}=A_{\mu} + \partial_{\mu} \Lambda$. Once the
Hamiltonian formalism of Maxwell theory is done, the configuration
variable is the 3-dimensional part of $A_{\mu}$, namely, $A_a$
while the temporal part of $A_{\mu}$, namely, $A_0$ becomes $-
\lambda$ with $\lambda$ the Lagrange multiplier associated with
the Gauss law. The initial Lagrangian gauge symmetry $A_{\mu}
\rightarrow {A'}_{\mu}=A_{\mu} +
\partial_{\mu} \Lambda$ is, in the canonical formalism, encoded in the gauge
transformation for the 3-dimensional connection $A_a \rightarrow {A'}_a = A_a
+ \partial_a \Lambda$ plus the transformation law for the Lagrange multiplier
$\lambda \rightarrow \lambda' = \lambda - {\dot \Lambda}$. In (anti-)
self-dual gravity, the Lagrangian action depends on the 4-dimensional
connection $A_{\mu}^i = \Gamma_{\mu}^i - \epsilon i \omega_{\mu {\hat 0}}\,^i
$, $\Gamma^i_{\mu} = -\frac{1}{2} \epsilon^i\,_{jk} \omega_{\mu}\,^{jk}$
valued in the (complex) algebra of $SU(2)$. Once the $3+1$ decomposition of
spacetime is done, the 3-dimensional part of $A^i_{\mu}$ becomes the
configuration variable $A^i_a$ and its temporal part $A^i_0$ becomes minus
the Lagrange multiplier $\lambda^i$ associated with the Gauss constraint.
Here, as in Maxwell theory, the initial gauge symmetry present in the (anti-)
self-dual action (\ref{SDII}): 4-dimensional diffeomorphism invariance plus
internal Lorentz transformations of the tetrad frame is, in the Hamiltonian
formalism, encoded in the transformation law for the phase space variables
and for the Lagrange multipliers \cite{Mon01}, in particular, when spacetime
${\cal M}=\Sigma \times R$ and $\Sigma$ has no boundary, the action
(\ref{split}) is fully gauge-invariant on the constraint surface under the
gauge symmetry generated by the first class constraints and the Lagrange
multipliers \cite{Mon01}. In this sense, in spite of the fact (anti-)
self-dual gravity has a quadratic in the momenta first class Hamiltonian
constraint, it resembles Yang-Mills and Maxwell theories which contain a
Gauss constraint which is linear and homogeneous in the momenta.

Finally, in spite of the simplicity of the constraints in terms of Ashtekar
variables, self-dual gravity describes complex general relativity. To recover
the real sector associated with real general relativity, extra conditions
must be imposed; the so-called {\it reality conditions}. Nevertheless, the
various proposals to incorporate the reality conditions have not worked in
general and the issue of how to get the real sector of self-dual gravity is
still an open problem. In \cite{HAMT96}, a novel proposal to recover the real
sector of self-dual gravity was proposed. There, reality conditions are
implemented as second class constraints which lead to the introduction of
Dirac brackets to handle them, but the price we pay for having included
reality conditions as second class constraints is a non-polynomial Dirac
bracket. To avoid the use of Dirac brackets, an alternative proposal was
developed in \cite{Mon99a,Mon99b}. There, second class class constraints are
transformed into first class ones, following the method of \cite{Amorim}. The
physical meaning of this formulation and its utility in the quantization of
gravity remain unclear.

\section{Self-dual gravity with topological terms}
The idea is to add topological terms to (anti-) self-dual gravity. To do
this, it is worth writing the action (\ref{action}) in a different fashion
\begin{eqnarray}
S & = & \int \left [ (\alpha_2 + i \alpha_1 )
{}^{(+)} (e^I \wedge e^J) \wedge {^{(+)} R}_{IJ} +
(\alpha_2 - i \alpha_1 )
{}^{(-)} (e^I \wedge e^J) \wedge {^{(-)} R}_{IJ}
\right ] \nonumber\\
& & + \int \left [
(\alpha_3 + i \alpha_4 ) {}^{(+)} R^{IJ} \wedge {^{(+)} R}_{IJ} +
(\alpha_3 - i \alpha_4 )
{}^{(-)} R^{IJ} \wedge {^{(-)} R}_{IJ}
\right ] \label{actionII}\, .
\end{eqnarray}
Note that when $\alpha_2 - i \alpha_1=0$ the second term in (\ref{actionII})
vanishes and the first one corresponds to the self-dual action. Therefore,
because just the self-dual part of the connection enters with this choice, it
is natural to consider $\alpha_3 - i \alpha_4  = 0$ and also drop the fourth
term which involves the anti-self-dual connection. With these choices, the
remaining action is just a functional of the self-dual connection and of the
tetrad frame
\begin{eqnarray}
S [e , {}^{(+)} \omega ] & = & \int \left [ 2 i \alpha_1 \,\,
{}^{(+)} (e^I \wedge e^J) \wedge {^{(+)} R}_{IJ} + 2 i \alpha_4
{}^{(+)} R^{IJ} \wedge {^{(+)} R}_{IJ} \right ] \label{dual} \, .
\end{eqnarray}
Last action is the self-dual action (\ref{SDII}) supplemented with the
self-dual
Pontrjagin term. The anti-self-dual analog of (\ref{dual}) is given with
$\alpha_2 + i \alpha_1 =0$ and $\alpha_3 + i \alpha_4  = 0$. In summary,
self-dual ($\epsilon=+$) and anti-self-dual ($\epsilon=-$) gravity with
Pontrjagin term are given by $\alpha_2 = \epsilon i \alpha_1$, and
$\alpha_3 = \epsilon i \alpha_4$. Thus, the action
\begin{eqnarray}
S [ e , \omega ] & = &
+ \alpha_1 \; \int \ast ( e^I \wedge e^J ) \wedge R_{IJ} (\omega)
+  \epsilon i \alpha_1  \; \int e^I \wedge e^J  \wedge R_{IJ} (\omega)
\nonumber\\
& & + \epsilon i \alpha_4 \int R^{IJ} (\omega) \wedge R_{IJ} (\omega)
+ \alpha_4 \int {\ast R}^{IJ} (\omega) \wedge R_{IJ} (\omega)
\, , \label{SDA}
\end{eqnarray}
which can be rewritten as
\begin{eqnarray}
S [e , {}^{(\epsilon)} \omega ] & = & \int \left [ 2 \epsilon i \alpha_1 \,\,
{}^{(\epsilon)} (e^I \wedge e^J) \wedge {^{(\epsilon)} R}_{IJ} + 2 \epsilon i
\alpha_4 {}^{(\epsilon)} R^{IJ} \wedge {^{(\epsilon)} R}_{IJ} \right ] \, ,
\label{SDAII}
\end{eqnarray}
is the right generalization for (anti-) self-dual gravity to include
topological terms. As already mentioned, both Euler and Pontrjagin terms in
(\ref{SDA}) combine to become a complex (self-dual or anti-self-dual)
Pontrjagin term in (\ref{SDAII}). Now it follows the canonical formalism of
(\ref{SDAII}). The `time gauge' is chosen, namely, $e^0= N dx^0$, $e^i =
E^i_a N^a dx^0 + E^i_a dx^a$ where $a=1,2,3$ denotes space indices. The
action (\ref{SDAII}) becomes
\begin{eqnarray}
S [A^i_a ,{\widetilde \pi}^a_i , \ut{\lambda} , \lambda^a ,
\lambda^i ] & = &
\int dx^0 dx^3 \left [ {\dot A}^i_a {\widetilde \pi}^a_i -
(\ut{\lambda} {\widetilde {\widetilde H}} +
\lambda^a {\widetilde V}_a + \lambda^i {\widetilde G}_i )
\right ] \nonumber\\
& & + \int dx^0 \int dx^3 \partial_a ({\widetilde \pi}^a_i \lambda^i) \, ,
\end{eqnarray}
where the dependence of the phase space variables and the Lagrange
multipliers on the initial Lagrangian variables is
\begin{eqnarray}
{\widetilde \pi}^a_i & := & {\widetilde \Pi}^a_i +
2 \epsilon i \alpha_4 {\tilde \eta}^{abc} F_{bci} =
-2 \epsilon i \alpha_1 E E^a_i +
2 \epsilon i \alpha_4 {\tilde \eta}^{abc} F_{bci}
\, , \nonumber\\
A^i_a & := & \Gamma_{a}^i - \epsilon i \omega_{a {\hat 0}}\,^i
\, , \quad \Gamma^i_a = -\frac{1}{2} \epsilon^i\,_{jk} \omega_a\,^{jk}
\, , \nonumber\\
\ut{\lambda} & := & -\frac{1}{4 \alpha_1}
\frac{N}{E} \, , \nonumber\\
\lambda^a & := & N^a \, , \nonumber\\
\lambda^i & := & - A^i_0 = - (\Gamma_0^i - \epsilon i
\omega_{0 {\hat 0}}\,^i )
\, , \quad \Gamma^i_0 = -\frac{1}{2} \epsilon^i\,_{jk} \omega_0\,^{jk} \, ,
\label{DEFII}
\end{eqnarray}
and the first class constraints have the form
\begin{eqnarray}
{\widetilde{\widetilde H}} & := & \epsilon^{ijk}
( {\widetilde \pi}^a_i - 2 \epsilon i \alpha_4
{\widetilde \eta}^{acd} F_{cd i} )
( {\widetilde \pi}^b_j - 2 \epsilon i \alpha_4
{\widetilde\eta}^{bef} F_{ef j} )
F_{abk} \ , \nonumber\\
{\widetilde V}_a & := & ({\widetilde \pi}^b_i - 2 \epsilon i \alpha_4
{\widetilde\eta}^{bcd} F_{cd i}) F_{ab}\,^i \ , \nonumber\\
{\widetilde G}_i & := & {\cal D}_a {\widetilde \pi}^a_i  \ .
\label{CII}
\end{eqnarray}
Thus, the effect of the complex Pontrjagin term added to the self-dual action
is as follows: i) a `magnetic' component $2 \epsilon i \alpha_4 {\tilde
\eta}^{abc} F_{bci}$ is added to the standard (anti-) self-dual momentum
${\widetilde \Pi}^a_i = -2 \epsilon i \alpha_1 E E^a_i$ [see Eq. (\ref{DEF})]
to obtain the new momentum of (\ref{DEFII}), ii) the expressions of the first
class constraints are modified too [see Eq. (\ref{CI}) and Eq. (\ref{CII})],
iii) finally, notice that even though $\alpha_1$ (and therefore $\alpha_2$)
is fixed an equal to $\alpha_1 = \frac{c^3}{16 \pi G}$, the parameter
$\alpha_4$ (and therefore $\alpha_3$) is, in principle, a free parameter.

In summary, the contribution in the canonical formalism of the term
proportional to the Bianchi identities when there is no torsion, which is
related to the Barbero-Immirzi parameter, is completely different to the
contribution coming from the topological terms. The former affects (when the
time gauge is chosen) to the configuration variable while the latter affects
its momentum. Therefore, Barbero-Immirzi ambiguity present in general
relativity can not be, directly, identified as coming from a topological
term, rather, the present analysis suggests it that the analogy between the
$\theta$-angle of Yang-Mills theory and the Barbero-Immirzi of gravity is not
appropriate. It would be worth performing the canonical analysis of action
(\ref{action}) in two more cases: a) without fixing the gauge. At first
sight, this would imply a bigger phase space with the corresponding
introduction of second class constraints to recover the correct counting of
degrees of freedom, b) without restricting to the (anti-) self-dual case,
i.e., leaving $\alpha_2$, $\alpha_3$, and $\alpha_4$ as arbitrary real
parameters. As the present paper suggests, this fact would imply phase space
variables $(A^i_a , {\widetilde \pi}^a_i)$ for the gravitational field, where
$A^i_a = \Gamma^i_a + \beta K^i_a$ and ${\widetilde \pi}^a_i = {\widetilde
\Pi}^a_i + \gamma {\widetilde \eta}^{abc} F_{bci}$; with $(A^i_a ,
{\widetilde \Pi}^a_i )$, ${\widetilde \Pi}^a_i= \frac{1}{\beta} {\widetilde
E}^a_i$ the canonical pair of Barbero's formulation, i.e., two free
parameters $\beta$ and $\gamma$ would appear in the formalism if local
Lorentz symmetry is destroyed. It might be interesting to perform a Lorentz
covariant canonical analysis of the action (\ref{action}) too.

Finally, some words on the implications of the topological terms in the
quantum theory. From the form of the constraints (\ref{CII}) it follows that
the quantum theory might feel the topological properties of spacetime.

\section*{Acknowledgements}
The author would like to thank an anonymous referee for his detailed
criticisms to the first version of this work. Also the author thanks
financial support from the {\it Sistema Nacional de Investigadores} (SNI) of
CONACyT.


\end{document}